\documentclass[12pt]{article}
\usepackage{amssymb}
\title{Finite width effects in the model of unstable particles
 with random mass}
\author{V.I.Kuksa}
\date{Institute of Physics, Rostov State University,\\
 pr. Stachki 194, Rostov-on-Don, 344090 Russia,\\
 E-mail address: kuksa@list.ru, ph.8632-202874}
\begin{document}

\maketitle

\begin{abstract}

A phenomenological model of unstable particles based on
uncertainty principle is discussed in quantum field approach. We
show that the simplest quantum field description of mass
uncertainty makes it possible to account finite width effects for
particles with large decay widths.

PACS number(s): 11.10St, 13.20Jf, 13.38Dg.

 \newpage

\end{abstract}

\pagenumbering{arabic} \setcounter{page}{1}

\section{Introduction}

To test the Standard Model predictions in processes with
participation of $W$ - , $Z$ - bosons and $t$ - quark one should
take into account finite width or instability effects \cite{RC96}.
At last time these effects have been evaluated in the cases of
$t\rightarrow WZb$ \cite{GA01} and some hadron decays \cite{HK00}
- \cite{VK03}. Such calculations strongly depend on the unstable
particle conception, which is the subject of intensive discussion
now.

Quantum field description of unstable particles (UP) has been
discussed during last five decades \cite{HL54} - \cite{MV63}. Two
directions of investigation were founded in these works --
"propagator" and "spectral function" approaches. The conventional
way to take into account the instability effects consist in the
Dyson resummation of self-energy in "propagator" approach
\cite{ML59} - \cite{MV63}. This procedure is the most direct and
consistent quantum field description of UP \cite{RJ61} -
\cite{WB88} and has got development in numerous works \cite{GB88}
- \cite{BK02}. However the realization of this program runs into
problems connected with the requirements of being unitary, gauge
invariance and with procedure of renormalization in perturbation
theory \cite{MV63, RS91, AA94, JP95, AB00, MN01}. Precise and
consistent definitions of mass and width for UP have been
notoriously difficult yet, caused by above mentioned problems. The
principal source of methodical difficulties is connected with the
fact that UP's have finite lifetime and therefore lie somewhat
outside the traditional formulation of quantum field theory
\cite{MV63, PG01}. As a result some phenomenological methods are
used in description of UP and resonance lines.

The second direction was formulated by Matthews and Salam in
\cite{PM58, PM59} and is connected with Lehmann (spectral)
representation \cite{HL54}. It is based on the uncertainty
principle for unstable quantum system, which leads to the
uncertainty relation for mass of unstable particle in the rest
frame \cite{PM58, ML59}:

\begin{equation}\label{E:1}
  \delta m\,\, \delta \tau \sim 1\,\,\,\longrightarrow\,\,\,\delta m
  \sim \Gamma_{tot}\,\,\,(c=\hslash=1).
\end{equation}

For UP with large $\Gamma_{tot}$ the large value of mass
uncertainty leads to noticeable modification of decay properties
or to, so called, "mass smearing" effects. The value of these
effects were calculated in phenomenological way for $B^0$ and
$\Lambda^0$ decays \cite{HK00}, \cite{AK92} - \cite{TU92'} and for
decay channel $t\rightarrow ZWb$ in convolution and decay-chain
method \cite{GA01}.

In this work the model realization of "mass smearing" idea (mass
uncertainty) is represented in the quantum field framework. The
main element of the proposed model is the simplest generalization
of field operator function which describes the UP as particles
with non-fixed masses. It was shown that the model is convenient
and simple tool for evaluation of finite width effects. The model
predictions are in agreement with experiment and describes some
peculiarities in generation and decay processes of particles with
large total widths. Short version of the model was represented in
\cite{VK03} where, in particular, large instability effects for
some hadron decays were discussed in details.

\section{The model of unstable particles with random mass
parameter}

In accordance with uncertainty principle the  model field operator
function is represented as superposition of ordinary ones weighted
by some model function $\omega(\mu)$. For simplicity we consider
scalar field:

  \begin{equation}\label{E:2}
    \varphi (x)=\int\omega(\mu) \varphi_{\mu}(x)\,d\mu ,
  \end{equation}
where $\varphi_{\mu} (x)$ is usual field operator, which describes
the state with fixed mass $m^2=\mu$:
  \begin{equation}\label{E:3}
    \varphi_{\mu}(x)=\frac{1}{(2\pi)^{3/2}}\int\varphi_{\mu}(k)
    \delta(k^{2}-\mu)e^{ikx}\,d^{4}k.
  \end{equation}

In (\ref{E:2}) and (\ref{E:3}) parameter $\mu$ has the status of
$m^2$ because $\mu=k^2$. For stable particle
$\omega(\mu)=\delta(\mu-M^2)$ and for UP $\omega(\mu)$ describes
the finite width or "mass smearing" effects. The expressions
(\ref{E:2}) and (\ref{E:3}) describes ensemble of unstable
particles and can be interpreted as a decision of motion equation
which follows from model Lagrangian for "free" fields:
  \begin{equation}\label{E:4}
    L(\varphi(x))=\frac{1}{2}\int |\omega(\mu)|^2(\partial_{k}\varphi_{\mu}(x) \partial^{k}\varphi_{\mu}(x)
      - \mu \,\varphi_{\mu}(x)\varphi_{\mu}(x))\,d\mu.
  \end{equation}
Thus, we discuss the approach where "spreading" caused by
interaction of UP with decay channels is described by wave packet
(\ref{E:2}) but corresponding field operator function has the
status of initial "free" field. Such approach, as it will be shown
in this section, is some phenomenological alternative to
propagator renormalization method.

 Commutation relations have an additional $\delta$ -
function of parameter $\mu$:

  \begin{equation}\label{E:5}
    [\varphi_{\mu}^{-}(\vec{k}) ,
    \dot{\varphi}_{\mu^{'}}^{+}(\vec{k}^{'})]_{-} =
    \delta(\vec{k}-\vec{k^{'}}) \delta(\mu-\mu^{'}),
  \end{equation}
where creation and annihilation operators are defined as
$\varphi^{+,-}_{\mu}(\vec{k})=\varphi^{+,-}_{\mu}(k)/\sqrt{2k^0}$
and $k^0=\sqrt{\vec{k}^2+\mu}$. Relation (\ref{E:5}) means
additional assumption: the acts of creations and annihilations of
particles with various $\mu$ don't interfere. So, the parameter
$\mu$ has status of physically distinguishable value as $m^2$. Now
we'll show that (\ref{E:2}) , (\ref{E:3}) and (\ref{E:5}) lead to
Lehmann type spectral representation of causal Green function. In
coordinate representation from (\ref{E:5}) and (\ref{E:3}) it
follows:
  \begin{equation}\label{E:6}
    [\varphi_{\mu}^{-}(x) , \dot{\varphi}_{\mu^{'}}^{+}(y)]_{-} =
    \delta(\mu-\mu^{'})\,
    \frac{1}{i}\, D_{\mu}^{-}(x-y).
  \end{equation}
In (\ref{E:6}) $D_{\mu}^{-}(x-y)$ is Pauli -- Jordan function
defined as:
  \begin{equation}\label{E:7}
    D_{\mu}^{-}(x-y) = \frac{i}{(2\pi)^{3}} \int\frac{d\vec{k}}{2k_{\mu}^{0}}\,\, e^{-ik(x-y)},
  \end{equation}
where $k_{\mu}^{0} = \sqrt{\vec{k}^{2}+\mu}$. The function
$D_{\mu}^{+}(x-y)$ is defined in analogy with (\ref{E:7}). Taking
into account (\ref{E:2}) we can get spectral representation of
Pauli -- Jordan function:
  \begin{equation}\label{E:8}
    [\varphi^{-}(x),\dot{\varphi}^{+}(y)]_{-} =\frac{1}{i} D^{-}(x-y)=
    \frac{1}{i} \int \rho(\mu)D_{\mu}^{-}(x-y)\,d\mu,
  \end{equation}
where $\rho(\mu)=|\omega(\mu)|^{2}$ is model probability density.

The causal Green function
  \begin{equation}\label{E:9}
    i\langle T[\varphi(x) \varphi(y)]\rangle_{0} =D^{C}(x-y),
  \end{equation}
can be expressed through the Pauli -- Jordan functions:
  \begin{equation}\label{E:10}
    D^{C}(x-y)=\theta(x^{0}-y^{0})D^{-}(x-y)-\theta(y^{0}-x^{0})D^{+}(x-y).
  \end{equation}
Using (\ref{E:8}) we get spectral representation of casual
function
  \begin{equation}\label{E:11}
    D^{C}(x)=\int \rho(\mu)D_{\mu}^{C}(x)\,d\mu ,
  \end{equation}
where:
  \begin{equation}\label{E:12}
    D_{\mu}^{C}(x)=\frac{1}{(2\pi)^{4}} \int \frac{e^{-ikx}}{\mu-k^{2}-i\varepsilon}\,\,.
  \end{equation}
In momentum representation from (\ref{E:11}) and (\ref{E:12}) it
follows
 \begin {equation}\label{E:13}
    D^{C}(k)=\int \frac{\rho(\mu)\,d\mu}{\mu-k^{2}-i\varepsilon}\,\,.
  \end{equation}
There is one undetermined yet element in the model --\, the
probability density $\rho(\mu) = |\omega(\mu)|^2$. To find
$\rho(\mu)$ we have identified the model Green function $D^C(k)$
with standard renormalized propagator by means of analytical
continuation to complex plane $k^2 \rightarrow z$:
 \begin{equation}\label{E:14}
  \frac{1}{z-M^2_0-\Sigma(z)}\,\,\,\longleftrightarrow\,\,\,
  \int^{\infty}_{\mu_0} d\mu \frac{\rho(\mu)}{z-\mu} \,\, = D^C(z),
 \end{equation}
where $\Sigma(k^2\pm i\epsilon) = Re\Sigma(k^2)\mp i
Im\Sigma(k^2)$ \cite{RJ61} and $\mu_0$ is threshold. From
(\ref{E:14}) and Cauchy theorem we have:
 \begin{equation}\label{E:15}
  D^C(\mu+i\epsilon) - D^C(\mu-i\epsilon) = \oint_{\Gamma} dz
  \frac{\rho(z)}{\mu-z} = -2\pi i \rho(\mu),
 \end{equation}
where $\Gamma$ is the contour with cut along real axis from
$\mu_0$ to positive infinity. On the other hand, right side of the
expression (\ref{E:15}) can be represented with help of
identification (\ref{E:14}) in the form:
 \begin{equation}\label{E:16}
  D^C(\mu+i\epsilon)-D^C(\mu-i\epsilon)=\frac{-2iIm\Sigma(\mu)}
  {[\mu-M^2_0-Re\Sigma(\mu)]^2+[Im\Sigma(\mu)]^2}\,.
 \end{equation}
From (\ref{E:15}) and (\ref{E:16}) one can easily get the
expression for probability density:
 \begin{equation}\label{E:17}
  \rho(\mu)=\frac{1}{\pi}\,\, \frac{Im\Sigma(\mu)}
   {[\mu-M^2_0-Re\Sigma(\mu)]^2+[Im\Sigma(\mu)]^2}\,.
 \end{equation}
From (\ref{E:17}) it follows that model weight function
$\omega(\mu)$ can be defined in the form:
 \begin{equation}\label{E:18}
  \omega(\mu)=\frac{1}{\sqrt{\pi}}\,\,
  \frac{\sqrt{Im\Sigma(\mu)}}{\mu-M^2_0-\Sigma(\mu)}
 \end{equation}
At peak range $\mu\approx M^2_0+Re \Sigma(M^2)$ we have usual
Breight--Wigner type (Lorentzian) distribution:
 \begin{equation}\label{E:19}
  \rho(\mu)\approx
  \frac{1}{\pi}\,\,\frac{M\Gamma}{(\mu-M^2)^2+M^2\Gamma^2}\,,
 \end{equation}
where $M^2=M^2_0+Re\Sigma(M^2)$ and $M\Gamma=Im\Sigma(M^2)$. The
identification (\ref{E:14}) establishes connection between
discussed model and "propagator" method.

\section{Generation and decay of unstable particles}

The state vector is determined in a standard way with additional
variable $\mu$:
 \begin{equation}\label{E:20}
    |\vec{k},\mu\rangle = \dot{\varphi}_{\mu}^{+}(\vec{k})
    |0\rangle ;\quad  \varphi_{\mu}^{-}(\vec{k})|0\rangle = 0.
  \end{equation}
Then the transition amplitude contains $\omega(\mu)$ as a result
of commuting
 \begin{equation}\label{E:21}
    [\varphi^{-}(x),\dot{\varphi}_{\mu}^{+}(\vec{k})]_{-}=\frac{1}{(2\pi)^{3/2}}
    \frac{\omega(\mu)}{\sqrt{2k_{\mu}^{0}}} \exp [-ikx] ,
 \end{equation}
where $\,k_{\mu}^{0}=\sqrt{\vec{k}^{2}+\mu}$ and
$\dot{\varphi}^+_{\mu}(\vec{k})$ -- creation generator of state
vector. From this result it follows that transition amplitude is a
product of weight function $\omega(\mu)$ and expression for
amplitude $A^{st}$, calculated in standard (conventional) way with
fixed mass $\mu=m^2$:
 \begin{equation}\label{E:22}
    A(k,\mu)=\omega(\mu) A^{st}(k,\mu).
 \end{equation}
In (\ref{E:22}) the value $k$ stands for all kinematics variables.
The expression for differential probability of transition is
 \begin{equation}\label{E:23}
     dP(k,\mu)=|\omega(\mu)|^2 |A^{st}(k,\mu)|^{2}\,d\mu ,
 \end{equation}
where $\rho(\mu)=|\omega(\mu)|^2$ is the model probability
density. From (\ref{E:23}) we can straightway get the formula for
decay width in form of standard one weighted by $\rho(\mu)$:
 \begin{equation}\label{E:24}
    \Gamma(\bar\mu,\sigma)=\int^{\mu_u}_{\mu_d} \rho(\mu;\bar\mu,\sigma)\,
    \Gamma^{st}(\mu)\,d\mu\,,
 \end{equation}
where $\mu_d$ and $\mu_{u}$ depend on threshold and total energy
of processes. In (\ref{E:24}) for simplicity we represent
$\rho(\mu)$ by some two-parametric distribution function
$\rho(\mu; \bar\mu,\sigma)$ with mean value $\bar\mu \approx M^2$
and mean square deviation $\sigma \approx \Gamma/2$. In general
case we should to determine $\rho(\mu)$ with help of (\ref{E:18})
or (\ref{E:19}) but for rough evaluation we can approximate
$\rho(\mu)$ by phenomenological two-parametric distribution
function. In this case our model reduces to the phenomenological
account of finite width effect. The expression (\ref{E:24}) can be
applied when UP with large total width is in both initial or final
state. Moreover, it can be easy generalized to the case when there
are two or more such particles. When $\rho(\mu) = \delta(\mu-M^2)$
we get usual result in fixed mass approach. In general case
(\ref{E:24}) leads to modification of phase space, threshold
"smearing" and $s$ -- dependence of width.

To illustrate the deviation of model predictions from conventional
ones we choose Gaussian approximation for $\rho(\mu)$ due to
suitable asymptotic behavior in infinity (large $\mu_{max}$ ). In
a case of heavy boson decay to two fermions when $M_f\ll M_V$ and
$\bar\mu=M^2_V,\, \sigma=\Gamma^{tot}_V/2$ we get
 \begin{equation}\label{E:25}
  \Gamma^M(\bar\mu,\sigma)/\Gamma^{st}(\bar\mu) \approx
  1+3\Gamma^2_{tot}/4M^2\,,
\end{equation}
where $\Gamma^M$ and $\Gamma^{st}$ are model and standard
predictions for discussed partial widths. Decay low is subjected
to analogous modification. For small time $t/\tau(\bar\mu)\ll1$ we
have deviation from exponential low:
 \begin{equation}\label{E:26}
   \frac{N(t)}{N_{0}}\approx 1-(1+\frac{3\Gamma_{tot}^{2}}{4M^{2}})\frac{t}{\tau(M)}.
 \end{equation}
It should be noted that the model modification of decay low
differs from one discussed in \cite{AP80} - \cite{EN88}. From
(\ref{E:25}), (\ref{E:26}) it follows that the model corrections
in the discussed case are rather small ($\sim\Gamma^2_{tot}/M^2$).
However in the cases of near threshold decays ($m_1+m_2\approx M$)
this corrections are very large (see the next section).

\section{Experimental test of the model}

Finite width effects in decays of fundamental UP with large
$\Gamma_{tot}$, such as $Z, W$ - bosons and $t$ - quark, should be
taken into account in precise measurements. The deviation of
$Br^M(Z\rightarrow f\bar{f})/Br^{st}(Z\rightarrow f\bar{f})$ from
unity according to (\ref{E:25}) is equal to
$3\Gamma^2_{tot}/4M^2_Z\approx6*10^{-4}$. So, the effect of
instability in such channels gives corrections an order of\,
$0.1\%$ that is an order of least experimental errors in $Z$ -
physics and much less than errors in $W$ - physics. We need more
precise both experimental data and theoretical calculations in
Standard Model.

Significantly more large effects of instability (finite width)
take place in near threshold processes when $(M_i-M_f)\sim
\Gamma_{tot}$, for example in decays $Z\rightarrow Wbc$ or
$t\rightarrow WZb$. Last process was discussed in detail in
\cite{GA01} without and with account of instability effects. These
effects were evaluated in the frame of so called "decay-chain"
method and "convolution" method where double weighting and two
distribution of invariant mass (for $W$ and $Z$) was applied in
formal analogy with our mass parameter distribution $\rho(\mu)$.
It was fond in \cite{GA01} that
 \begin{center}
   $10^{-6}<Br(t\rightarrow WZb)<10^{-5}$\,\,\,for
   $(170<m_t<180)\, Gev,$
 \end{center}
while usual calculation with fixed masses gives us
$Br(t\rightarrow WZb)=0$ for $M_t<M_W+M_Z+M_ b$ and
$Br(t\rightarrow WZb)=10^{-7}$ for $M_t=180\, Gev$. Unfortunately,
even result obtained with account of widths effects makes the
observation of this decay channel at LHC very difficult.

Instability effect more accessible for observation can occur in
decay channel $Z\rightarrow Wbc$ and should be taken into
consideration in precision $Z$ - physics. The value of the model
correction to $Br(Z\rightarrow Wbc)$ mainly caused by modification
of phase space. For discussed process the phase space $R(M_i)$ in
nonrelativistic case can be expressed in the form \cite{EB73}:
 \begin{equation}\label{E:27}
  R(M_i)\approx \frac{\pi^3}{2}\,\,\frac{\sqrt{M_W M_b
  M_c}}{(M_W+M_b+M_c)^{3/2}}\,\,(M_Z-M_W-M_b-M_c)^2.
 \end{equation}
Using double Gaussian (for simplicity) weighting with parameters
$\bar{\mu}_Z=M^2_Z$, $\sigma_Z=\Gamma^{tot}_Z/2$ and
$\bar{\mu}_W=M^2_W$, $\sigma_W=\Gamma^{tot}_W/2$ from (\ref{E:27})
we get:
 \begin{equation}\label{E:28}
  \frac{Br^M(Z\rightarrow Wbc)}{Br^{st}(Z\rightarrow Wbc)}\approx
  1+\frac{\sigma^2_Z+\sigma^2_W}{(M_Z-M_W-M_b-M_c)^2} \approx 1.1
 \end{equation}
So, the rough model evaluation of correction caused by finite
width effect in rare decay channel $Z\rightarrow Wbc$ gives the
value $\approx 10\%$. This correction should be taken into account
when precision measurements are compared with theoretical
prediction.

The effects of "mass smearing" have large value in the processes
of generations and decays of hadrons with large total widths.
Hadrons are not fundamental particles and quantum field approach
can't be applied in general case. But "mass smearing" effect
follows from fundamental uncertainty principle and takes place at
various hierarchy levels. Proposed model does not describe hadron
decays but gives us a simple way to evaluate instability effects
as correction to traditional calculations.

The first punctual evaluations of finite width effects in heavy
hadron decays were fulfilled in \cite{HK00}, \cite{AK92} -
\cite{TU92'}. The phenomenological Breight-Wigner type weighting
of expressions for widths was applied in these works for decays
$B^0\rightarrow D^-\rho^+, D^-a^+_1;\,\, B^0_s\rightarrow
D^-a^+_1$ and $\Lambda^0_b\rightarrow \Lambda^+_c a^-_1$. The
results of calculations reveal that contributions of "mass
smearing" effects are large -- from 20\% to 40\%\,, and its
accounting improves considerably the conformity of experimental
data and theoretical predictions.

One of the most pure effect of "mass smearing" in hadron physics
takes place in decay channels $\phi(1020)\rightarrow K^+K^-,
K_LK_S$. The ratio of branchings does not depends on hadron
factors in good approximation and is equal to the ratio of phase
space \cite{EF02}:
 \begin{equation}\label{E:29}
  k=\frac{Br(\phi\rightarrow K^+K^-)}{Br(\phi\rightarrow K_LK_S)}=
  \frac{g^2_+}{g^2_0}(\frac{1-4m^2_+/m^2_{\phi}}{1-4m^2_0/m^2_{\phi}})^{3/2}\,.
 \end{equation}
There is a discrepancy between experimental and theoretical values
of $k$ when $g^2_+ = g^2_0$, which was discussed in \cite{EF02}:
 \begin{center}
  $k^{exp}=1.456\pm 0.033,\,\,\,k^{th}=1.528$
 \end{center}
Various corrections to $k^{th}$ have been evaluated in \cite{AB00}
but discrepancy has increased only (Fermi "gold rule" puzzle). The
model prediction $k^M$ depends on $\mu_{max}$ and in
Breight-Wigner type approximation for $\rho(\mu)$:
 \begin{center}
  $k^M=1.42 - 1.49$\, when $\mu_{max}=(1 - 3)\, Gev,$\,\,
  $k^M=k^{exp}$\, when $\mu_{max}=2\, Gev.$
 \end{center}
So, the model can resolve discussed problem with help of
reasonable assumption concern $\mu_{max}$. Analogous result was
received in \cite{EF02} with assumption of phase space
s-dependence, which is similar to convolution method \cite{GA01}.

The decay channel $f_o(980)\rightarrow K\bar{K}$ is the example of
"mass forbidden" one ($M_{f_o} < 2M_K$) and we have the effect of
"threshold smearing". Model prediction for the ratio of forbidden
and dominant branchings ($g_K\sim g_{\pi}$):
 \begin{center}
  $Br(f_0\rightarrow K\bar{K})/Br(f_0\rightarrow 2\pi) \sim
  0.1$
 \end{center}
This rough estimation should be considered as the conformity of
theoretical prediction and experimental indication of channel
$Br(f_0\rightarrow K\bar{K})$ ("seen", \cite{PDG02}).

There are many examples of hadron decays with large total width
and of near threshold decay channels. The determination of
hadron's probability density $\rho(\mu)$ in quantum field approach
is limited due to its composite structure. However, as it had been
shown in \cite{VK03}, calculation of instability effects can be
done for some hadron decays with high accuracy (2 - 3 \%). This
problem needs more detailed consideration with help of
phenomenological methods.

\section{Conclusion}
Proposed model of UP is the simplest phenomenological realization
of uncertainty principle in the framework of quantum field theory.
The model calculation of decay rates is in formal analogy with the
"convolution" type treatment but model structure contains
phenomenological elements at more fundamental level. Quantum field
approach restricts application of the model to hadron decays but
successfully describes some peculiarities connected with finite
width effects. The model does not contradict to experimental data
on decays of fundamental particles and is in quantitative
agreement with the data on some hadron decay channels.

The principal element of discussed model is the wave packet
(\ref{E:2}) which describes initial or final "free" state vector
with commutation relations (\ref{E:5}). This packet is the result
of model accounting of interaction connected with decay channels,
which leads to "spreading" of mass. All information about this
interaction enter to the probability density $\rho(\mu)$. The
status of random mass parameter $\mu$ is determined by dispersion
condition $\mu=k^2$ and relations (\ref{E:5}) as physical random
mass squared of UP. This interpretation arises the question on
physical meaning of unstable particle in real and virtual states.


\begin{thebibliography}{99}
  \bibitem{RC96}
  Report CERN 96-01, Physics at LEP2, edited by G. Altarelli, T.
  Sjostrand, F.Zwirner.
  \bibitem{GA01}
  G. Altarelli, L. Conti, V. Lubicz, Phys. Lett. B 502 (2001) 125.
  \bibitem{HK00}
  H. Kaur and M.P. Khanna, J. Phys. C 26 (2000) 387.
  \bibitem{EF02}
  E. Fischbach, A.W. Overhauser, B. Woodahl, Phys. Lett. B 526
  (2002) 355.
  \bibitem{VK03}
  V.I. Kuksa, Semyphenomenological model of unstable particles,
  in: Proc. 17th International Workshop on High Energy Physics and
  Quantum Field Theory. September, 2003, Samara, Russia (in press).
  \bibitem{HL54}
  H. Lehmann, Nuovo Cimento, 11 (1954) 342.
  \bibitem{HA57}
  H. Araki, Y. Munakata et al., Progr. Theor. Phys. 17 (1957) 419.
  \bibitem{PM58}
  P.T. Matthews and A. Salam, Phys. Rev. 112 (1958) 283.
  \bibitem{PM59}
  P.T. Matthews and A. Salam, Phys. Rev. 115 (1959) 1079.
  \bibitem{ML59}
  M. Levy, Nuovo Cimento, 13 (1959) 115.
  \bibitem{JS60}
  J.Schwinger, Ann.Phys. 9 (1960) 169.
  \bibitem{RJ61}
  R. Jacob and R.G. Sachs, Phys. Rev. 121 (1961) 350.
  \bibitem{MV63}
  M. Veltman, Physica, 29 (1963) 186.
  \bibitem{WB88}
  W. Beenakker and W. Hollik, Z. Phys. C 40 (1988) 141.
  \bibitem{GB88}
  G. Burgers, CERN Sci. Rept. 6/1 (1988) 121.
  \bibitem{SW91}
  S. Willenbrock and G. Valencia, Phys. Lett. B 259 (1991) 373.
  \bibitem{AL91}
  A. Leike, T. Riemann and J. Rose, Phys. Lett. B 273 (1991) 513.
  \bibitem{RS91}
  R.G. Stuart, Phys. Lett. B 272 (1991) 353.
  \bibitem{RS93}
  R.G. Stuart, Phys. Rev. Lett. 70 (1993) 3193.
  \bibitem{AA94}
  A. Aeppli, G.J. Oldenborgh, D. Wyler, Nucl. Phys. B 428 (1994) 126.
  \bibitem{HV94}
  H. Veltman, Z. Phys. C 62 (1994) 35.
  \bibitem{JP95}
  J. Papavassiliou, A. Pilaftsis, Phys. Rev. Lett. 75 (1995) 3060.
  \bibitem{LM98}
  L. Maiani and M. Testa, Ann.Phys. 263 (1998) 353.
  \bibitem{MP98}
  M. Passera and A.Sirlin, Phys. Rev. D 58 (1998) 113010.
  \bibitem{DB99}
  D. Bardin and G. Passarino, The Standard Model in the Making,
  Oxford University Press, 1999.
  \bibitem{AB00}
  A.R. Bohm, N.L. Harshman, Nucl. Phys. B 581 (2000) 91.
  \bibitem{AB001}
  A.R. Bohm, N.L. Harshman et al., Eur. Phys. J. C 18 (20000) 333.
  \bibitem{PG01}
  P.A. Grassi, B.A. Kniehl and A.Sirlin, Phys. Rev. Lett. 86 (2001) 389.
  \bibitem{MN01}
  M.L. Nekrasov, Eur. Phys. J. C 19 (2001) 441.
  \bibitem{BK02}
  B.A. Kniehl, A. Sirlin, Phys. Lett. B 530 (2002) 129.
  \bibitem{AK92}
  A.N. Kamal and R.C. Verma, Phys. Rev. D 45 (1992) 982.
  \bibitem{TU92}
  T. Uppal and R.C. Verma, Z. Phys. C 56 (1992) 273.
  \bibitem{TU92'}
  T. Uppal, R.C. Verma, Phys. Rev. D 46 (1992) 2982.
  \bibitem{AP80}
  A. Peres, Ann. Phys. 129 (1980) 33.
  \bibitem{LK82}
  L.A. Khalfin, Phys. Lett. B 112 (1982) 223.
  \bibitem{CC82}
  C.B. Chiu, B. Misra and E.C.G. Sudarshan, Phys. Lett. B 117 (1982) 34.
  \bibitem{EN88}
  E.B. Norman, S.B. Gazes et al., Phys. Rev. Lett. 60 (1988) 2246.
  \bibitem{EB73}
  E. Bycling, K. Kajantie, Particle Kinematics, London, 1973.
  \bibitem{AB00}
  A. Bramon, R. Escribano et al., Phys. Lett. B 486 (2000) 406.
  \bibitem{PDG02}
  Particle Data Group, K.Hagiwara et al., Phys. Rev. D 66 (2002) 010001.
  \end{thebibliography}
\end{document}